\documentclass[11pt]{article}
\usepackage{hyperref}
\usepackage{tikz} 

\usepackage{kpfonts}

\usepackage{tikz-cd}

\usepackage{amsmath,amsthm,amsfonts,amssymb,amsopn,amscd,bm} 
\usepackage{color}

\DeclareMathOperator{\Tr}{Tr}

\def\qed{{\unskip\nobreak\hfil\penalty50
\hskip2em\hbox{}\nobreak\hfil$\square$
\parfillskip=0pt \finalhyphendemerits=0\par}\medskip}
\def\proof{\trivlist \item[\hskip \labelsep{\bf Proof.\ }]}
\def\endproof{\null\hfill\qed\endtrivlist\noindent}

\def\tilde{\widetilde}

\def\Ad{{\hbox{\rm Ad}}}

\def\a{\alpha}

\def\e{\varepsilon}
\def\g{\gamma}

\def\l{\lambda}

\def\A{{\cal A}}
\def\B{{\cal B}}

\def\C{{\cal C}}
\def\D{{\cal D}}
\def\E{{\cal E}}
\def\F{{\cal F}}

\def\M{{\cal M}}

\def\R{{\cal R}}

\def\H{{\cal H}}

\def\S{{\cal S}}

\def\f{{\varphi}}
\def\s{{\sigma}}

\def\l{{\lambda}}
\def\x{{{h}}}

\def\PSL{{{\rm PSL}(2,\mathbb R)}}

\def\S2{S^{1(2)}}

\def\RR{\mathbb R}

\newtheorem{theorem}{Theorem}[section]
\newtheorem{lemma}[theorem]{Lemma}

\newtheorem{corollary}[theorem]{Corollary}

\newtheorem{proposition}[theorem]{Proposition}

\theoremstyle{remark} 

\newcommand{\ben}{\begin{equation}}
\newcommand{\een}{\end{equation}}

\newcommand{\bthm}{\begin{theorem}}
\newcommand{\ethm}{\end{theorem}}

\newcommand{\bprop}{\begin{proposition}}
\newcommand{\eprop}{\end{proposition}}
\newcommand{\bcor}{\begin{corollary}}
\newcommand{\ecor}{\end{corollary}}
\newcommand{\blem}{\begin{lemma}}
\newcommand{\elem}{\end{lemma}}

\def\PSL{PSU(1,1)}

\def\CC{{\mathbb C}}

\def\SL2{{{\rm SL}(2,\R)}}

\def\PSL2{{{\rm PSL}(2,\Reali)}}

\def\U1{{{\rm V}(1)}}
\def\SU2{{{\rm SV}(2)}}

\def\SU{{{\rm SU}}}

\def\A{{\mathcal A}}
\def\B{{\mathcal B}}
\def\C{{\mathcal C}}
\def\D{{\mathcal D}}
\def\F{{\mathcal F}}
\def\H{{\mathcal H}}

\def\M{{\mathcal M}}

\def\P{{\mathcal P}}

\def\S{{\mathcal S}}

\def\RR{{\mathbb R}}

\def\a{\alpha}

\def\g{\gamma}

\def\l{\lambda}

\def\x{\xi}

\def\s{\sigma}

\def\f{\varphi}

\title{\Huge A note on continuous entropy}

\author{{\sc Roberto Longo
}
\\
Dipartimento di Matematica,
Universit\`a di Roma ``Tor Vergata'',\\
Via della Ricerca Scientifica, 1, I-00133 Roma, Italy\\
E-mail:  longo@mat.uniroma2.it
\\
\phantom{X}\\
{\sc Edward Witten}
\\
Institute for Advanced Study,
School of Natural Sciences,\\
Einstein Drive, Princeton, NJ 08540\\
E-mail: witten@ias.edu}

\date{}

\begin{document} 

\maketitle

\begin{abstract}
Von Neumann entropy has a natural extension to the case of an arbitrary semifinite von Neumann algebra, as was considered by I. E.  Segal.  We relate this entropy to the relative entropy and show that the entropy increase for an inclusion of von Neumann factors is bounded by the logarithm of the Jones index. The bound is optimal if the factors are infinite dimensional. 
\end{abstract}

\newpage

\section{Introduction}

In recent years, there has been much interest in applications of entropy to quantum field theory.   For example, a novel argument \cite{CasiniHuerta} for   the Zamolodchikov $c$-theorem concerning the irreversibility of renormalization group flow in 2 spacetime dimensions made use of the formal notion of the von Neumann entropy of the density matrix of a quantum field reduced to a local region in spacetime, a double cone.   It is difficult to put this argument on a rigorous basis because in quantum field theory the algebra of a double cone region is actually a von Neumann algebra of Type III, and notions such as density matrix and von Neumann entropy are not available for such  algebras.  

In the physics literature, the non-existence of a notion of entropy for an algebra of Type III is described by saying that in quantum field theory, the entropy of a double cone region (or of any local region in spacetime) is ultraviolet divergent. The nature of the ultraviolet divergence depends on the spacetime dimension. In two dimensions, the divergence is only logarithmic. The argument in \cite{CasiniHuerta} relies on this and involves considering linear combinations of entropies for different double cone regions from which the divergences cancel.   Defining rigorously finite linear combination of entropies might be one way to put the argument in \cite{CasiniHuerta}, and others somewhat like it, on a rigorous basis.   This would require considerations beyond the von Neumann algebra structure, since the assertion that the ultraviolet divergence of the entropy is logarithmic is special to 2 spacetime dimensions.

Here we will consider a somewhat similar but much simpler situation in which a renormalized notion of entropy is available.    
Entropy for an algebra $\A$ of Type II$_1$ was first discussed long ago by I. E. Segal \cite{S}.   Segal noted that for a state of $\A$, a fairly natural notion of entropy can be defined, with the unusual property that there is a (normalized) state of $\A$ with maximum entropy, namely the tracial state $\tau$, and no state  of minimum entropy.    Entropy is then defined to vanish for the tracial state, and therefore, to be negative (or equal to $-\infty$ in some cases) for  other states.  
To see the interpretation of this entropy in terms of renormalization, consider a hyperfinite Type II$_1$ algebra $\A$, which is the large $i$ limit of a family of matrix algebras $\M_i$ of dimension $n_i^2$.  The maximum entropy state of $\M_i$ is the tracial state $\tau_i$, 
with von Neumann entropy $S_{\mathrm{vN}}(\tau_i)=\log n_i$. In the limit $i\to\infty$, $\tau_i$ converges to $\tau$.   If 
$\f_i$ is a family of states of $\M_i$ that converge for large $i$ to a state $\f$ of $\A$,  the Type II$_1$ entropy $S(\f)$ 
can be defined as  $\lim_{i\to\infty} (S_{\mathrm{vN}}(\f_i)-S_{\mathrm{vN}}(\tau_i))=\lim_{i\to\infty} (S_{\mathrm{vN}}(\f_i)-\log n_i)$.  With this definition, it is clear that $\tau$ is the maximum entropy state of $\A$ and has entropy 0.   The subtraction of the divergent part $\log n_i$ is necessary to ensure the existence of a large $i$ limit, and makes clear the analogy with renormalized entropy as studied in the physics literature.

A renormalized entropy can also be defined for an algebra $\A$ of Type II$_\infty$, with the difference that in this case, because there is no canonical normalization of a tracial weight of $\A$,   entropy is really only naturally defined  up to an additive constant, the same for all states.
At first sight, one might think that because the algebras of local regions in quantum field theory are of Type III, entropy of a Type II algebra would have little application in physics.   In fact, this thought is probably one reason that entropy of an algebra of Type II has been relatively little-studied.
However, Type II algebras can appear in black hole physics \cite{W} and also in certain random matrix models \cite{V}, so there is indeed some physical motivation to study entropy for Type II.

In this article, we observe that entropy for a state $\f$ of a Type II algebra can be interpreted as a relative entropy $S(\f |\!| \tau)$.
This implies that in a trace-preserving inclusion of Type II algebras $\B\subset \A$, the entropy of any state increases, cf. \cite{S}.  Our main result is a bound on the entropy increase in terms of the Jones index $[\A:\B]$.

Vaughan Jones made fundamental contributions in multiple areas of mathematics and mathematical physics.  He always was extremely interested in applications of operator algebras to physics.    The Jones index of a subfactor was an important tool in his discoveries in von Neumann algebra theory, mathematical physics, and knot theory.   We hope therefore that the modest contribution to the theory of the Jones index that we make here would have pleased Vaughan, and we are happy to dedicate this article to his memory.

\section{Noncommutative probability spaces}
Let $(X,\mu)$ be a probability space and $f\in L^1(X,\mu)$, with $f> 0$ almost everywhere, and $\int_Xfd\mu = 1$. 
Define the entropy $S(f)$ of the random variable $f$ by
\[
S(f) = -\int_X f\log fd\mu = \int_X f\log f^{-1} d\mu \, .
\]
\blem
$S(f)\leq 0$. 
\elem
\proof
Note that $-\log$ is a convex function, thus
\[
-\log\Big(\int_X f^{-1} d\nu \Big) \leq  -\int_X \log f^{-1} d\nu
\]
by Jensen inequality, for every probability measure $\nu$ on $X$. Therefore, setting $d\nu = fd\mu$, we have
\ben\label{SN}
S(f) = \int_X f\log f^{-1}d\mu = \int_X \log f^{-1}d\nu \leq \log\Big(\int_X f ^{-1}d\nu \Big) = \log\Big(\int_X 1 d\mu \Big)
=  0\, .
\een
\endproof
Note that
\[
S(f) = -S(\nu |\!| \mu)
\]
where $S(\nu |\!| \mu)$ is the relative entropy of the between the states $\mu$ and $\nu$ on the von Neumann algebra $L^\infty(X,\mu)$, indeed
\[
S(\nu |\!| \mu) = \int (\log f - \log 1) d\nu = \int f\log f d\mu\, .
\]
Let now $\A$ be a finite von Neumann algebra, thus there exists a normal faithful trace $\tau$ ($\tau$ is unique if $\A$ is a factor once $\tau$ is normalised with $\tau(1) =1$). If $\f$ is a normal faithful state of $\A$, there exists a positive, non singular operator $\rho$ affiliated with $\A$ such that 
\ben\label{E1}
\f (x) = \tau(\rho x)\, , \quad x\in\A\, ,
\een
thus $\tau(\rho) =1$.  

We define the {\it entropy} $S_\tau(\f)$ of $\f$ w.r.t. $\tau$ as
\[
S_\tau(\f) = -\tau(\rho\log\rho)\, .
\]
Setting $\tau_0 = \frac1{\tau(1)} \tau$, we have $\f = \tau_0(\rho_0\cdot)$ with $\rho_0 = \tau(1) \rho$, therefore
\[
S_{\tau_0}(\f) = -\tau_0(\rho_0\log\rho_0)= S_\tau(\rho) - \tau(1)\log\tau(1)
\]
so we can assume that $\tau$ is normalised with $\tau(1) = 1$. 
\bprop
If $\tau$ is normalised, we have
\[
S_\tau(\f) \leq 0 \, ,
\]
possibly $S_\tau(\f) = -\infty$, and $S_\tau(\f) = 0$ iff $\f = \tau$. 
\eprop
\proof
By considering the von Neumann algebra generated by $\rho$, the statement follows by \eqref{SN}. Alternatively, the statement is a consequence of the following proposition. 
\endproof
Let $\A$ be an arbitrary von Neumann algebra and $\f,\psi$ normal, positive, faithful linear functionals on $\A$. 
 Araki's relative entropy \cite{Ar76} is defined by
\ben\label{AE} 
S(\f |\!| \psi) = -(\xi_\f , \log\Delta_{\x_\psi , \xi_\f}\xi_\f)\, ,
\een
where $\xi_\f,\xi_\psi$ are any cyclic vector representatives of $\f,\psi$ on the underlying Hilbert space $\H$ (we may assume that $\A$ is in a standard form). 
Here $\Delta_{\xi_\f,\xi_\psi}$ is the relative modular operator between $\f$ and $\psi$ \cite{T}, i.e. $\Delta_{\xi_\f,\xi_\psi}={\rm S}^*_{\xi_\f,\xi_\psi}{\rm S}_{\xi_\f,\xi_\psi}$ with ${\rm S}_{\xi_\f,\xi_\psi}$ the closure of the map $x\xi_\psi\mapsto x^*\xi_\f$, $x\in\A$. The right hand side of \eqref{AE} is well defined for all $\f, \psi$ by 
\[
S(\f |\!| \psi) = \int_0^\infty \log s\, d(\xi_\f, e_s \xi_\f) \,
\]
with $\Delta_{\x_\psi , \xi_\f}= \int_0^\infty s\, de_s$ the spectral resolution of $\Delta_{\x_\psi , \xi_\f}$. Indeed
\[
S(\f |\!| \psi) \geq \f(1)\big( \log\f(1) - \log \psi(1)\big) \, ,
\]
possibly $S(\f |\!| \psi) = +\infty$. Recall also that, if $\f$ is a state, then
\ben\label{norm}
S(\f |\!| \l \psi) = S(\f |\!| \psi) - \log\l \, ,\quad \l>0\, .
\een
Note that $S(\f |\!|\psi)$ can be easily defined also if $\f,\psi$ are not faithful, see \cite{OP}; for simplicity we mostly consider the faithful case. 
\bprop\label{SS}
Let $\tau$ be a normal faithful trace on $\A$. Then
\[
S_\tau(\f) = -S(\f |\!| \tau)\, ,
\]
where $S(\f |\!| \tau)$ is Araki's relative entropy \eqref{AE} between $\tau$ and $\f$ on $\A$. 
\eprop 
\proof
The relative modular operator $\Delta_{\xi_\tau\xi_\f}$ (w.r.t. vector representatives $\xi_\f,\xi_\tau$ of $\f,\tau$ in the natural cone) is equal to $\rho^{-1}$. Therefore
\[
S(\f |\!| \tau) = - (\xi_\f, \log \Delta_{\xi_\tau,\xi_\f}\xi_\f) = (\xi_\f, \log \rho\, \xi_\f) = \f(\log \rho) = \tau(\rho\log \rho) = -S_\tau(\f)\, .
\]
\endproof
Due to the above proposition, the entropy $S_\tau$ depends on the choice of the tracial state $\tau$. However, if $\A$ is a type II$_1$ factor, the tracial state is unique. 

Since the relative entropy is monotone, we infer that the $S_\tau$ is monotone. 
\bcor
If $\B \subset\A$ is a von Neumann subalgebra and $\tau$ a normal faithful trace on $\A$.  Then
\[
S_\tau(\f|_\B) \geq S_\tau(\f) \, ,
\]
where $S_\tau(\f|_\B)$ is the entropy of the restriction of $\f$ to $\B$ w.r.t. $\tau|_\B$. 
\ecor
\proof
\[
S_\tau(\f|_\B) = -S(\f|_\B |\!| \tau|_\B) \geq -S(\f |\!| \tau) = S_\tau(\f) \, .
\]
\endproof
We have the additivity of $S_\tau$.
\bprop
Let $\A_i$ be von Neumann algebras with tracial normal faithful states $\tau_i$, $i=1,2$, and $\tau = \tau_1\otimes\tau_2$ the trace on $\A = \A_1\otimes\A_2$. Then
\[
S_\tau(\f_1\otimes \f_2) = S_{\tau_1}(\f_1)  + S_{\tau_2}(\f_2)
\]
for any normal faithful states $\f_i$ on $\A_i$. 
\eprop
\proof
\[
S_\tau(\f_1\otimes \f_2) = S(\f_1\otimes \f_2 |\!| \tau_1\otimes\tau_2) = S(\f_1 |\!| \tau_1) + S(\f_2 |\!| \tau_2) 
= S_{\tau_1}(\f_1)  + S_{\tau_2}(\f_2)\, .
\]
\endproof
Suppose now $\M_n$ is type $I_n$ factor, namely $\M_n$ is the $n\times n$ matrix algebra. Let $\Tr$ be the trace on $\M_n$, thus $\Tr(1) = n$ and 
\[
\Tr = n\tau\, ,
\]
with $\tau$ the normalised trace. With $\f$ a state on $\M_n$, the von Neumann entropy of $\f$ is defined by
\[
S_{\rm vN}(\f) = -\Tr(\s\log\s)\, ,
\]
with $\s$ the density matrix associated with $\f$, namely
\ben\label{E2}
\f (x) = \Tr(\s x)\, , \quad x\in\M_n\, ,
\een 
Note that the von Neumann entropy on type I factor is not monotone. 

We now compare $S_\tau(\f)$ with $S_{\rm vN}(\f)$. With $\rho$ and $\s$ as in \eqref{E1} and \eqref{E2}, clearly we have
\[
n\s = \rho
\]
thus the following holds as particular case of \eqref{norm}. 
\bprop\label{tvN}
If $\tau$ is normalised, we have
\[
S_\tau(\f)   = S_{\rm vN}(\f) -  \log n \, .
\]
\eprop
\proof
We have
\begin{multline*}
S_{\rm vN}(\f) = -\Tr(\s\log\s) = -n\tau \Big(\frac{\rho}{n}\log\frac{\rho}{n}\Big) 
= -\tau \Big(\rho\log\frac{\rho}{n}\Big) \\ = -\tau (\rho\log\rho) + \log n = S_\tau(\f) +  \log n \, .
\end{multline*}
\endproof
As a consequence, the entropy $S_\tau(\f) $ on $\M_n$ satisfies the bound
\ben\label{Si}
-\log n \leq S_\tau(\f) \leq 0\, ;
\een
$S_\tau(\f)  = 0$ iff $\f$ is the tracial state $\tau$, while $S_\tau(\f) = - \log n$ iff $\f$ is a pure state. 

Now, the von Neumann entropy is additive: 
\[
S_{\rm vN}(\f_1\otimes \f_2) = S_{\rm vN}(\f_1)  + S_{\rm vN}(\f_2)
\]
therefore, if $\f_1$ is a state on $\A$ and $\f_2$ is a state on $\M_n$, we have
\[
S_\tau(\f_1\otimes \f_2) = S_\tau(\f_1)  + S_\tau(\f_2) = S_\tau(\f_1)  +  S_{\rm vN}(\f_2) -  \log n \, .
\]
Let  $\A_1, \A_2$ be von Neumann algebras $\f$ a normal faithful state on $\A = \A_1\otimes \A_2$ and $\psi_k$ a faithful normal state on $\A_k$, $k=1,2$. Recall the subadditivity property of the relative entropy \cite[Cor. 5.21]{OP}:
\[
S(\f |\!| \psi_1 \otimes \psi_2) \geq S( \f|_{\A_1} |\!| \psi_1) + S( \f|_{\A_2} |\!| \psi_2) \, . 
\]
\bprop
Let $\A$ be a finite von Neumann algebra with faithful normal tracial state $\tau$. If $\f$ is a faithful state on $\A\otimes\M_n$, then
\[
S_\tau(\f|_\A) + S_{\rm vN}(\f|_{\M_n}) - S_\tau(\f) \geq \log n\, .
\]
The equality occurs iff $\f = \f|_\A \otimes \f|_{\M_n}$. 
\eprop
\proof
We have
\begin{multline*}
S_\tau(\f) = - S(\f|\!|\tau) = - S(\f|\!|\tau_\A\otimes \tau_{\M_n}) \leq
- S(\f|_\A|\!|\tau_\A) - S(\f|_{\M_n}|\!| \tau_{\M_n})=\\
S_\tau(\f|_\A) + S_\tau(\f|_{\M_n}) = S_\tau(\f|_\A) + S_{\rm vN}(\f|_{\M_n}) - \log n\, .
\end{multline*}
\endproof
\section{A bound for the entropy increase}
Let $\B\subset \A$ be an inclusion of von Neumann algebras and let $\tau$ be a faithful normal tracial state on $\A$. In case $\A$, $\B$ are factors, V. Jones defined the index $[\A:\B]$ as the ratio of Murray and von Neumann's coupling constants. In the more general non factor case,
Pimsner and Popa \cite{PP} gave a probabilistic definition of the index $[\A:\B]_\e$, with $\e: \A \to \B$ the trace preserving expectation:
\ben\label{PPi}
[\A:\B]_\e = \l^{-1}\, ,
\een
with $\l \geq 0$ the best constant such that
\ben\label{PP}
\e(x) \geq \l x
\een
for all positive $x\in\A$. The above inequality is called the Pimsner-Popa inequality. 

If $\A$ and $\B$ are II$_1$ factors and $\e$ is the unique trace preserving conditional expectation $\e:\A\to\B$, then
\[
[\A:\B]_\e = \text{Jones index of $\B\subset\A$}\, .
\]
We shall later comment on the case $\A$ is not of type II$_1$. 

Let now $\B\subset \A$ be an arbitrary inclusion of von Neumann algebras with a normal faithful conditional expectation $\e:\A\to\B$.  
Given  normal states $\f$ and $\psi$ on $\A$, recall the formula \cite[Thm. 5.15]{OP} for the relative entropy
\ben\label{Petz}
S(\f |\!| \psi\cdot\e) = S(\f|_\B |\!| \psi|_\B) + S(\f |\!| \f\cdot\e) \, .
\een
\blem\label{boundla}
Let $\B\subset \A$ be an inclusion of von Neumann algebras and $\e:\A\to\B$ a finite index normal conditional expectation. Then
\[
S(\f |\!| \f\cdot\e) \leq  \log [\A:\B]_\e 
\]
for every faithful normal state of $\A$. 
\elem
\proof
As $\e(x) \geq \l x$, for all positive $x\in\A$, with $\l$ the inverse of the index, we have
\[
\f\cdot \e\geq \l \f 
\]
and this implies
\ben\label{el}
S(\f |\!| \f\cdot\e) \leq S(\f |\!| \l \f) = -\log\l \, ,
\een
where the first inequality follows by Corollary \ref{Sprop} and the equality $S(\f |\!| \l \f) = -\log\l$ is a particular case of \eqref{norm}. 
\endproof
\bprop\label{bound}
Let $\B\subset \A$ be an inclusion of von Neumann algebras, $\tau$ a finite normal faithful trace on $\A$ and $\e: \A \to \B$ the trace preserving expectation. For every normal faithful state $\f$ on $\A$, 
we have
\ben\label{boundf}
S_{\tau|_\B}(\f|_\B) -S_\tau(\f)  \leq \log[\A:\B]_\e \, ,
\een
where $[\A:\B]_\e$ is the index w.r.t. $\e$. 
\eprop
\proof
Taking $\psi = \tau$ in formula \eqref{Petz}, we have
\ben\label{Se}
S(\f |\!| \tau) = S(\f|_\B |\!| \tau|_\B) + S(\f |\!| \f\cdot\e) \, ,
\een
namely
\[
S_{\tau|_\B}(\f|_\B) - S_\tau(\f) =  S(\f |\!| \f\cdot\e) \, .
\]
As by Lemma \ref{boundla}
\[
S(\f |\!| \f\cdot\e) \leq S(\f |\!| \l \f) = -\log\l = \log[\A:\B]_\e\, ,
\]
the bound \eqref{boundf} follows.
\endproof
It is known that if $\A$ is infinite-dimensional, then the quantity $[\A:\B]_\e$ as we have defined is
equal to the Jones index $[\A:\B]$.   For finite-dimensional $\A$, this is actually not the case.   For example, if
$\A=\M_n$, $\B=\CC$, then $[\A:\B]=n^2$ but $[\A:\B]_\e =n$ .    
The definition of $[\A:\B]_\e$ can be modified as follows to coincide with the Jones index $[\A:\B]$ in all cases:  $[\A:\B]$ is the inverse of the largest constant $\l$ such that $\e - \l\!\cdot\!{\rm id}$ is completely positive.   However, except in the case that $\A$ is finite-dimensional, $\e - \l\!\cdot\!{\rm id}$ is completely positive if and only if it is positive, and this refinement is unnecessary.

We conclude by providing a bound as in Prop. \ref{bound} for the increase of $S(\tau|\!| \f)$. 
Note that, with $\A$, $\tau$, $\f$ as in \eqref{E1}, we have
\[
S(\tau|\!| \f) =  -\tau(\log \rho) \geq 0 \, ,
\]
similarly as in \eqref{SN}. 
\bprop\label{Bf}
With the notations in Prop. \ref{bound}, we hav
\[
S(\tau|\!| \f) -S(\tau|_\B |\!| \f|_\B)   \leq   \log[\A:\B]_\e \, .
\]
\eprop
\proof
Clearly
\[
\f|_\B = \tau\big(\e(\rho)\,\cdot\,\big)\, ,
\]
and $\e(\rho) \geq \l \rho$ by the Pimsner-Popa inequality.  So 
\[
\log\big(\e(\rho)\big) \geq \log(\l\rho)   = \log\rho + \log\l 
\]
because the logarithm is an operator monotone function. Therefore
\[
S(\tau|\!| \f) - S(\tau|_\B |\!| \f|_\B)   =  -\tau\big(\log(\e(\rho)\big) + \tau(\log\rho) \leq - \log\l  = \log[\A:\B]_\e \, .
\]
\endproof
It would be to interesting estimate $\sup_\f S(\f\cdot \e |\!| \f)$ too.

Note that, by the argument in the proof of Theorem \ref{bound2}, if $\A_i$ and $\B_i$ are increasing sequences of matrix subalgebras of dimension $n_i^2$ and $m_i^2$ such that $\cup_i \A_i$, $\cup_i \B_i$ are weakly dense in $\A$, $\B$, then 
\[
S_{\tau|_{\B}}(\f|_{\B})  - S_{\tau}(\f) = 
\lim_i\big(S_{\tau_i|_{\B_i}}(\f_i|_{\B_i})  - S_{\tau_i}(\f_i) \big)
= \lim_i\big(S_{\rm vN}(\f_i|_{\B_i}) - S_{\rm vN}(\f_i) + \log (n_i/m_i)\big)\, .
\]
Furthermore, if $\B_i\subset\A_i$ and $\e_{\B_{i+1}} \e_{\A_i} = \e_{\B_i}$, where $\e_{\A_i}$, $\e_{\B_i}$ denote the trace preserving expectations onto $\A_i$, $\B_i$, then by \cite[Prop. 2.6]{PP} we have
\[
[\A:\B]_\e = \lim_i [\A_i: \B_i] = \lim_i n^2_i/m^2_i \, ,
\]
therefore
\[
S_{\tau|_{\B}}(\f|_{\B})  - S_{\tau}(\f) =  \lim_i\big(S_{\rm vN}(\f_i|_{\B_i}) - S_{\rm vN}(\f_i)\big) +\frac12 \log [\A:\B]_\e \, .
\]

\section{State/weight relative entropy}
Let $\A$ be a von Neumann algebra and $\f,\psi$ positive, normal, linear functionals on $\A$. We recall Kosaki's variational formula \cite{KoJOT}. Fix any $^*$-strongly dense linear subspace $V$ of $\A$ containing the identity. Then
\ben\label{Ko}
S(\f |\!|\psi) = \sup_{n\in\mathbb N}\sup_{x\in\mathfrak V}\left\{ \f(1)\log n - 
\int_{1/n}^\infty\big(\f(y(t)^*y(t)) + t^{-1}\psi(x(t)x(t)^*)\Big)\frac{dt}{t}\right\} \, ,
\een
where $\mathfrak V$ is the set of all step functions $x: (1/n, \infty)\to V$ with finite range, and $x(t) + y(t) =1$. 
The advantage of Kosaki's formula is that it has all main properties built in it. 

Let now $\f$ be a positive, normal, faithful linear functional on $\A$ and $\psi$ a normal, faithful, semifinite weight on $\A$, see \cite[Chapter VII]{T} (semifinite means that the definition domain of $\psi$ is weakly dense in $\A$).  We may assume that $\A$ acts standardly on the GNS Hilbert space $\H_\f$ of $\f$. Let $\f'$ be the normal, faithful, positive linear functional on the commutant $\A'$ of $\A$ given by 
\ben\label{fprime}
\f' =  (\xi_\f, \cdot\, \x_\f) \, ,
\een
where $\xi_\f\in\H_\f$ is the GNS vector. Let $d\psi/d\f'$ be Connes' spatial derivative between $\psi$ and $\f'$ \cite{C80}.

We define the {\it relative entropy between $\f$ and $\psi$} by
\ben\label{Sw}
S(\f |\!| \psi) = - (\xi_\f, \log (d\psi/d\f')\, \xi_\f)\, ,
\een
provided the above formula is well defined; this is the case, in particular, if $\xi_\f$ belongs to the domain of $\log (d\psi/d\f')$. More generally, let $d\psi/d\f' = \int_0^\infty s\, d{e_s}$ be the spectral resolution of $d\psi/d\f'$, then 
\ben\label{Sspth}
S(\f |\!| \psi) =  -\int_0^\infty \log s\, d(\xi_\f , e_s\xi_\f)
\een
provided either the positive or the negative part of $\log s$ belongs to $L^1(\RR_+, d(\xi_\f , e_s\xi_\f))$. 
If $S(\f |\!| \psi)$ is well defined, then $S(\f |\!| \psi)$ can take any real value or $S(\f |\!| \psi) = \pm \infty$. We shall say that $S(\f |\!| \psi)$ is {\it finite} if $S(\f |\!| \psi)$ is well defined and $S(\f |\!| \psi)\neq\pm\infty$. 

If $\psi$ is bounded, then $d\psi/d\f'$ is equal to the relative modular operator $\Delta_{\xi_\psi,\xi_\f}$, where $\xi_\psi$ is a cyclic vector representative of $\psi$ in $\H_\f$, so $S(\f |\!| \psi)$ is Araki's relative entropy \eqref{AE} and is well defined for every normal faithful state $\f$. 

If $\psi=\tau$ is tracial, it follows, similarly as in Proposition \ref{SS}, that $d\tau/d\f' = \rho^{-1}$ with $\rho$ the density matrix of $\f$ as in \eqref{Ei}, so
\ben\label{Stau}
S(\f |\!| \tau) = - S_\tau(\f) = \tau(\rho\log\rho)
\een
and $S(\f |\!| \tau)$ is finite iff $\tau(|\rho\log\rho|)<\infty$ (see Section \ref{S4}). 

In particular, if $\A$ is a type I factor and $\tau$ is the usual trace $\Tr$ on $\A$, we have
\[
S(\f |\!| \Tr) = - \ \text{von Neumann entropy of $\f$}\, .
\]
Note that $S(\f |\!| \psi)$ can be defined also if the weight $\psi$ is not semifinite by restricting both $\f$ and $\psi$ to the weak closure of the definition domain of $\psi$. Still \eqref{Ko} holds. 

In the following, we shall use following elementary integral formula for the logarithm function:
\ben\label{log}
-\log\l = \int_0^\infty \big( (t+1)^{-1} -\l (t+\l)^{-1}\big)\frac{dt}{t}\, ,\quad \l>0\, .  
\een
\blem\label{Sin}
Let $\A$ be a von Neumann algebra, $\f$ a positive, normal, faithful linear functional on $\A$ and $\psi_1,\psi_2$ normal, faithful, semifinite weights on $\A$. If $S(\f |\!| \psi_1) $ and  $S(\f |\!| \psi_2)$ are well defined, 
then
\[
\psi_1 \leq \psi_2 \implies S(\f |\!| \psi_1)  \geq S(\f |\!| \psi_2) \, .
\]
\elem
\proof
We have $\psi_1 \leq \psi_2 \implies d\psi_1/d\f' \leq d\psi_2/d\f'$ \cite[Prop. 3.10]{T}. On the other hand, $d\psi_k/d\f' = (d\f' /d\psi_k)^{-1}$, so
\[
\psi_1 \leq \psi_2 \implies d\f' /d\psi_1 \geq d\f' /d\psi_2 \implies \log(d\f' /d\psi_1) \geq \log(d\f' /d\psi_2)
\]
because the logarithm is an operator monotone function. The right hand inequality means that $(\xi, \log(d\f' /d\psi_1 \xi) \geq (\xi, \log(d\f' /d\psi_2 \xi)$ for all $\xi$ in the common domain of $ \log(d\f' /d\psi_1)$ and $ \log(d\f' /d\psi_2)$ and follows by  \eqref{log}. So we have
\[
\psi_1 \leq \psi_2 \implies - (\xi_\f, \log (d\psi_1/d\f')\, \xi_\f) \geq - (\xi_\f, \log (d\psi_2/d\f')\, \xi_\f)
\]
if $\xi_\f$ is in the common domain. The more general case follows by the spectral theorem. 
\endproof
We shall say that $\psi$ has a {\it bounded entropy approximation} w.r.t. $\f$ if $S(\f |\!| \psi)$ is well defined and there exists a sequence of  positive, normal, faithful linear functionals on $\A$ such that $\psi_k(x) \nearrow\psi(x)$ for every positive $x\in\A$, and $S(\f |\!| \psi_k)$ is finite for some $k$, hence for all larger $k$. 
\blem\label{approx}
Let $\A$ be a von Neumann algebra, $\f$ a positive, normal, faithful linear functional on $\A$ and $\psi$ a normal, faithful, semifinite weight on $\A$. 
If $\psi$ has a bounded entropy approximation w.r.t. $\f$ with the $\psi_k$'s as above, then 
$
S(\f |\!| \psi_k) \searrow S(\f |\!| \psi)$.
\elem
\proof
By \cite[Cor. 3.13]{T}, we have $d\psi_k/d\f' \searrow d\psi/d\f'$. 
By formula \eqref{log} and Lebesgue monotone convergence theorem, we then have
$-(\xi_\f,  \log (d\psi_k/d\f') \xi_\f,)  \searrow -(\xi_\f, \log (d\psi/d\f') \xi_\f)$, where the expectation values are understood by the spectral theorem as in eq. \eqref{Sspth}.  So the Lemma is proved. 
\endproof
\bcor\label{Sprop}
Let $\A$ be a von Neumann algebra, $\f$ a normal, faithful, positive linear functional on $\A$ and $\psi$ a normal, faithful, semifinite weight on $\A$ with bounded entropy approximation w.r.t. $\f$. 

If $\B$ is a von Neumann algebra and $\a: \B \to \A$ a completely positive, normal, faithful, unital map such that $\psi\cdot\a$ is semifinite. Then $S(\f\cdot\a |\!| \psi\cdot\a)\leq  S(\f |\!| \psi)$ . 
\ecor
\proof
Let $\psi_k$ be a bounded entropy approximation sequence as above. For $k$ large enough,
\[
S(\f\cdot\a |\!| \psi\cdot\a)\leq 
S(\f\cdot\a |\!| \psi_k\cdot\a)\leq  S(\f |\!| \psi_k )\, ,
\]
where the first inequality also means that $S(\f\cdot\a |\!| \psi\cdot\a)$ is well defined, and follows by Lemma \ref{Sin}.  The second inequality follows by Kosaki's , see also \cite{Wit}. Then the corollary is a consequence of Lemma \ref{approx} by letting $k\to\infty$. 
\endproof
\blem\label{finiteS}
Let $\A$ be a von Neumann algebra, $\f,\psi$ faithful normal positive linear functional on $\A$ with $S(\f |\!|\psi )<\infty$ and $\e:\A\to\B$ a normal faithful conditional expectation. If $\f(1) =1$ and $\psi\cdot\e=\psi$, we have
\[
S(\f |\!| \psi) -  S(\f|_\B |\!| \psi|_\B) \leq  \log [\A:\B]_\e\, .
\]   
\elem
\proof
Of course, we may assume that $ [\A:\B]_\e <\infty$. 
By eq. \eqref{norm}, we may also assume that $\psi(1) =1$.  
We then immediately get
\[
S(\f |\!|\psi ) - S(\f|_\B |\!| \psi|_\B) =  S(\f |\!| \f\cdot\e)  \leq  \log [\A:\B]_\e\, ,
\]
where the equality is given by \eqref{Petz} and the inequality by Lemma \ref{boundla}. 
\endproof

\section{The bound in the semifinite case}\label{S4}
Let  $\A$ be a von Neumann algebra and $\tau$ a normal, faithful, semifinite trace on $\A$. With $\f$ a faithful, normal state on $\A$, there exists a positive, non-singular, selfadjoint operator $\rho$ affiliated to $\A$ (density matrix) such that $\f = \tau(\rho\, \cdot)$; namely $\f(x) = \tau(\rho^{1/2} x\rho^{1/2})$ for all positive $x\in\A$. 
The entropy $S_\tau(\f)$ is defined by
\ben\label{Ei}
S_\tau(\f) = -\tau(\rho\log\rho)\, ,
\een
provided $\tau(x)$ is well defined with $x\equiv\rho\log\rho$, namely either $\tau(x_+) < \infty$ or $-\tau(x_-) < \infty$, where $x_\pm$ is the positive/negative part of $x$. So $S_\tau(\f)$ is not defined for every normal state $\f$. We shall say that 
$S_\tau(f)$ is finite if both $\tau(x_+)$ and $\tau(x_-)$ are finite, namely $\tau(|\rho\log\rho|) <\infty$. 

Note that, even if $\A$ is a type II$_\infty$ factor, $S_\tau$ depends on the choice of the trace $\tau$, as the trace is unique only up to rescaling. However, the difference of entropies between two states is independent of the chosen trace $\tau$, due to the relation
\ben\label{resc}
S_{\l\tau}(\f)  = S_\tau(\f)  + \log\l\, ,\quad \l>0\, .
\een
The case $\tau$ is unbounded shows important differences with the case $\tau$ is bounded and
the notion of entropy $S_\tau(\f)$ needs care.  If $\A = L^\infty(\RR, dt)$ and $\tau$ is the Lebesgue integral, the state $\f$ is given by the integral with a positive density function $f\in L^1(\RR, dt)$ and
\[
 \int f\log f dt =  - S_\tau(\f) 
\]
is the {\it differential entropy} of $f$ introduced by Shannon. The differential entropy is neither positive nor negative definite. Moreover, it is not the limit of the discrete entropy under a discrete approximation, indeed one needs a logarithmic rescaling, see \cite[Chapter 8]{CT}. 
\blem\label{approxtau}
Let $\A$ be a von Neumann algebra, $\tau$ a normal, faithful, semifinite trace on $\A$ and $\f$ a normal, faithful, positive linear functional on $\A$ with finite entropy \eqref{Ei}.  Then $\tau$ has a bounded entropy approximation w.r.t. $\f$. 
\elem
\proof
Let $\rho$ be the density matrix of $\f$ w.r.t. $\tau$. By assumptions $\tau(\rho) < \infty$, $\tau(|\rho\log\rho|) < \infty$. Let $g_k$ be a sequence of positive Borel functions on $(0,\infty)$ such that $g_k\nearrow 1$ pointwise and $\tau(|\rho_k\log\rho_k|) < \infty$, with $\rho_k = \rho g_k(\rho)$. 
Thus
$\tau(\rho_k) < \infty$, 
$\rho_k\nearrow \rho$. 
With $\psi_k = \tau(\rho_k\, \cdot)$,  by the relation \eqref{Stau} the $\psi_k$'s give a bounded entropy approximation for $S(\f |\!| \tau)$. 
\endproof
As shown in \cite{S}, $S_\tau$ is monotone, provided the entropies are finite.
\bprop
Let $\B\subset\A$ be an inclusion of von Neumann algebras and $\tau$ a normal, faithful, semifinite, trace $\tau$ on $\A$ such that $\tau|_\B$ is semifinite. If $S_\tau(\f)$ is finite, then $S_{\tau|_\B}(\f|_\B)$ is well defined and
\[
S_{\tau|_\B}(\f|_\B) - S_\tau(\f)  \geq 0\,   .
\]
\eprop
\proof
By Cor. \ref{Sprop}, $
S(\f |\!| \tau) - S(\f|_\B |\!| \tau|_\B)  \geq 0$,
so the statement follows by eq. \eqref{Stau}. 
\endproof
We shall show that the bound by the logarithm of the index \eqref{boundf} still holds in the II$_\infty$ case, by extending the arguments in the previous section. Clearly, the entropy increase $S_{\tau|_\B}(\f|_\B) -S_\tau(\f)$ is independent of rescaling of $\tau$ due to \eqref{resc}. 

Now, let $\e:\A\to \B$ be a faithful normal conditional expectation. 
A definition of the index $[\A:\B]_\e$ for arbitrary inclusions of factors is given by the spatial theory \cite{Ko}, or by the crossed product \cite{L89}, and agrees with the Jones index in the II$_1$ case with $\e$ the trace preserving expectation. If $\A$ is a type III factor and
$\tilde\B\subset\tilde\A$ is the crossed product inclusion of von Neumann algebras in Takesaki's duality, then the index $[\A:\B]_\e$  shows up as the trace scaling factor 
\ben\label{tg}
\tau\cdot\g = [\A:\B]_\e \tau
\een
with $\g: \tilde\A\to \tilde\B$ the canonical endomorphism and $\tau$ the canonical trace on $\tilde \A$ \cite{L89}. 

The Pimsner-Popa inequality still holds, cf. \cite{L90}. Indeed $\l = [\A:\B]_\e^{-1}$ is the best constant such that $\e - \l\cdot{\rm id}$ is (completely) positive. 
In the non factor case, $[\A:\B]_\e$ is defined as the inverse of the best constant in the Pimsner-Popa inequality. 
\blem\label{fi}
Let $\B\subset\A$ be an inclusion of von Neumann algebras and $\tau$ a normal, faithful, semifinite, trace $\tau$ on $\A$ such that $\tau|_\B$ is semifinite.  There exists a type I subfactor $\F\subset \B$ and a tensor decomposition
\[
\B = \B_0\otimes \F \subset \A_0\otimes \F = \A
\]
such that $\tau = \tau_0\otimes \Tr$, with $\tau_0$ a tracial state on $\A_1$ and $\Tr$ the usual trace on $\F$. If $\e:\A\to \B$ is the trace preserving expectation, then we have a corresponding tensor decomposition of $\e$
\[
\e =\e_0\otimes{\rm id}\, ,
\]
with $\e_0 :\A_0\to \B_0$ preserving $\tau_0$, and
\[
[\A:\B]_\e =[\A_0:\B_0]_{\e_0}\, .
\]
\elem
\proof
The Lemma is essentially Proposition 2.3 of \cite{L89}. 
\endproof
\bthm\label{bound2}
Let $\A$ be a von Neumann algebra with a normal, faithful, semifinite trace $\tau$, $\B\subset\A$ a von Neumann subalgebra such that $\tau|_\B$ is semifinite and $\e: \A \to \B$ the trace preserving expectation. If $\f$ is a normal faithful state on $\A$ such that $S_\tau(\f)$ is finite, then 
$S_\tau(\f|_\B)$ is finite too and we have
\ben\label{bTh}
S_\tau(\f|_\B) -S_\tau(\f)  \leq \log[\A:\B]_\e \, ,
\een
where $[\A:\B]_\e$ is the index w.r.t. $\e$ and $S_\tau(\f|_\B) \equiv S_{\tau|_\B}(\f|_\B)$.  
\ethm
\proof
Fix the state $\f$. 
By Lemma \ref{approxtau}, there exists a sequence of positive linear functionals $\psi_k$ that give a bounded entropy approximation for $\tau$ w.r.t. $\f$. 
Thus $S(\f |\!| \psi_k) \searrow S(\f |\!| \psi)$ and $S(\f |\!| \psi_k)$ is finite for large $k$; so $S(\f|_\B |\!| \psi_k|_\B)$ is finite too for large $k$ because the relative entropy is monotone. 

Now, $\psi_k(x) \nearrow \tau(x)$ for all positive $x\in \A$, therefore $\psi_k(\e(x)) \nearrow \tau(\e(x)) =\tau(x)$ for all positive $x\in \A$. Moreover,
$S(\f |\!| \psi_k\cdot\e)$ is finite for large $k$ by formula \eqref{Petz} and Lemma \ref{boundla}. So we can assume that $\psi_k = \psi_k\cdot\e$. 

By Lemma \ref{finiteS} we have $S(\f |\!| \psi_k) -  S(\f|_\B |\!| \psi_k|_\B) \leq  \log [\A:\B]_\e$, therefore by Lemma \ref{approx} we get
\[
S(\f |\!| \tau) - S(\f|_\B |\!| \tau|_\B) =  \lim_k \big( S(\f |\!| \psi_k) - S(\f|_\B |\!| \psi_k|_\B) \big)  \leq \log[\A:\B]_\e
\]
and the proof is complete due to the relation \eqref{Stau}. 
\endproof
We end up this section by pointing out that the entropy of a state in a semifinite factor $\A$ depends only on the approximate inner equivalence. Namely,
\[
\f_1 \sim \f_2 \implies S_\tau(\f_1) = S_\tau(\f_2)\, ,
\]
where $\f_1 \sim \f_2$ means that the norm closed orbit by inner automorphisms in the predual $\A_*$ of $\A$ generated by $\f_1$ and $\f_2$ are the same; namely $\f_2$ belongs to the norm closure of $\{\f_1\cdot \Ad u : u \ \text{unitary of}\ \A\}$, where $\Ad u$ denotes the inner automorphism of $\A$ implemented by the unitary $u\in\A$. 
This follows because $\f_1 \sim \f_2$ iff 
the trace spectral density on the spectral family of the density matrices $\rho_i$ of $\f_i$ coincide, $i = 1,2$, \cite[Lemma 4.3]{HS}. Thus,
in this case, if $\rho_i = \int_0^\infty \l d e_{i,\l}$ is the spectral resolution of $\f_i$, we have $\tau(e_{1,\l}) = \tau(e_{2,\l})$ so that
\[
-\tau(\rho_i\log\rho_i) = -\int_0^\infty \l \log\l\, d \tau(e_{i,\l}) 
\]
is independent of $i$ (assuming the entropy is well defined). 

\section{The optimal bound}\label{optimal}
We now show that the bound given by Theorem \ref{bound2} is optimal for inclusions of infinite dimensional factors.  

Let $\B\subset\A$ be an inclusion of factors and $\e:\A\to\B$ be a normal faithful expectation. The index of $\B\subset\A$ w.r.t. $\e$ is finite if Haagerup's dual operator valued weight (\cite{Ha}, see \cite{T}) is a bounded map $\e^{-1}:\B'\to\A'$. Then $\e^{-1}$ is a scalar multiple of a conditional expectation $\e':\B'\to\A'$ and Kosaki's definition of the index \cite{Ko} is given by
\ben\label{Ki}
\e^{-1} = [\A:\B]_\e \e' \, .
\een
Unless $\A$ is finite dimensional, the index $ [\A:\B]_\e$ defined in \eqref{Ki} coincides with the index defined by the inequality \eqref{PP}, so we do not use a different symbol and specify the meaning of $ [\A:\B]_\e$ if necessary. 

In the following, we assume that $[\A:\B]_\e$ is finite and that $\A$ acts standardly on a Hilbert space $\H$. Let $\f$ be a faithful normal state
on $\A$ and $\xi_\f\in\H$ be a cyclic vector in $\H$ such that $\f = (\xi_\f,\cdot\,\xi_\f)$ on $\A$.  Denote by $\f'$ the state on $\B'$ given by $\f'=(\xi_\f,\cdot\,\xi_\f)$. 

The following relation has been derived by F. Xu in \cite[Prop. 2.4]{X}:
\[
S_\A(\f |\!| \f\cdot \e) + S_{\B'}(\f'|\!| \f'\cdot\e^{-1}) = 0\, ,
\]
therefore by \eqref{norm}
\ben\label{SSI}
S_\A(\f |\!| \f\cdot \e) + S_{\B'}(\f'|\!| \f'\cdot\e') = \log  [\A:\B]_\e \, ,
\een
where $S_\A$, $S_{\B'}$ denote the relative entropy in $\A$, $\B'$ and $ [\A:\B]_\e$ is the index in \eqref{Ki}. Here, the involved states are normal but not necessarily faithful, the relative entropies are defined, for example, by Kosaki's formula. 

The identity \eqref{SSI} is closely related to the functorial normalisation of the modular Hamiltonian in \cite{L18}.   
In the finite dimensional case, it has been discussed in \cite{MP}. 

Recall that a von Neumann algebra $\A$ is $\s$-finite iff it admits a faithful normal state; this is the case if $\A$ acts on a separable Hilbert space. 
For simplicity, the von Neumann algebras in this sections are $\s$-finite. 

With $\B\subset\A$ be an inclusion of factors on a Hilbert space $\H$, we call $\A'\subset\B'$ the dual inclusion on $\H$. 
\blem\label{dual}
A finite index inclusion of factors $\C\subset\D$ is isomorphic to the dual $\A'\subset\B'$ of an inclusion of factors $\B\subset\A$, with $\A$ acting standardly on a Hilbert space (namely there exists a cyclic and separating vector for $\A$), iff either $\D$ is infinite dimensional or ${\rm dim}(\C)^2/{\rm dim}(\D)$ is an integer.  
\elem
\proof
Suppose first that $\D$ is of type II$_1$. There exists a Jones' tunnel subalgebra $\E \subset \C$ for $\C\subset\D$, namely $\E$ is a subfactor of $\C$ such that $\E\subset\C\subset\D$ is Jones' extension \cite{Jo}. 
Let $\C$ act standardly on a Hilbert space $\H$, and let $\E\subset\C\subset \C_1$ be the Jones' extension of $\E \subset \C$ on $\H$. Then $\C\subset \D$ is isomorphic to $\C\subset \C_1$. On the other hand, $\C\subset \C_1$ is dual of $\C'_1\subset \C'$ and $\C'$ acts standardly on $\H$. 
So our lemma is proved in this case. 

The case $\D$ is an infinite factor is similar; in this situation, $\E = \g(\D)$ with $\g:\D\to\C$ the canonical endomorphisms \cite{L89}.

If $\D$ is finite dimensional, namely $\D$ is a matrix algebra, it is easy to see that a tunnel subalgebra $\E\subset \C$ for $\C\subset\D$ is a matrix subalgebra $\E\subset\C$ such that ${\rm dim}(\D)/{\rm dim}(\C) = {\rm dim}(\C)/{\rm dim}(\E)$. 
Since $\D$ is isomorphic to the tensor product $\C\otimes(\C'\cap\D)$, 
a tunnel subalgebra $\E\subset \C$ for $\C\subset\D$ exists iff $\C$ contains a subalgebra isomorphic to $\C'\cap\D$, namely
iff ${\rm dim}(\C'\cap\D) = {\rm dim}(\D)/{\rm dim}(\C)$ divides ${\rm dim}(\C)$, that is iff
${\rm dim}(\D)$ divides ${\rm dim}(\C)^2$. The rest of the finite dimensional proof is as in the type II$_1$ case. 
\endproof
Note that the condition that ${\rm dim}(\C)^2/{\rm dim}(\D)$ is an integer in Lemma \ref{dual} implies that the inverse of the Jones index $[\D:\C]$ w.r.t. the trace is the best constant in the Pimsner-Popa inequality \eqref{PP} for trace preserving expectation; indeed this holds iff ${\rm dim}(\C)^2\geq{\rm dim}(\D)$ \cite[6.5 Examples]{PP}. 

By the above lemma, both next Prop. \ref{SSP} and Cor. \ref{corO} remain true if $\A$ is finite dimensional and ${\rm dim}(\B)^2/ {\rm dim}(\A)$ is an integer. We state them in the infinite dimensional case for simplicity. 
\bprop\label{SSP}
Let $\B\subset \A$ be an inclusion of infinite dimensional factors and $\e:\A\to\B$ an expectation with finite index. Then 
\ben\label{max}
\log  [\A:\B]_\e  = \sup_\f S_\A(\f |\!| \f\cdot \e) \, ,
\een
where the supremum is taken over all normal states $\f$ of $\A$. 
\eprop
\proof
The inequality $S_\A(\f |\!| \f\cdot \e)\leq \log  [\A:\B]_\e$ has been shown in Lemma \ref{boundla}, and it also follows from \eqref{SSI}.  
The proposition is going to be proved by using eq. \eqref{SSI}. 

We may assume that $\A$ is in a standard form. 
We choose a faithful normal state $\f$ on $\A$ such that $\f\cdot\e =\f$; therefore $S_\A(\f |\!| \f\cdot \e)=0$. 
With $\xi_\f$ and $\f'$ as above,  eq. \eqref{SSI} gives 
\[
S_{\B'}(\f' |\!| \f'\cdot \e') = \log  [\A:\B]_\e
\] 
showing that the bound $\log  [\A:\B]_\e = \log  [\A':\B']_{\e'}$ in \eqref{max} is optimal for the dual inclusion $\A' \subset \B'$ with dual expectation $\e'$. So the proposition follows by Lemma \ref{dual}.  
\endproof
In order eq. \eqref{SSI} to hold, $\A$ was assumed to be in standard form. 
If $\A$ is finite dimensional, \eqref{max} does not hold in general with $ [\A:\B]_\e$ defined  in \eqref{Ki} according to Kosaki. Indeed,  we have $S_\A(\f |\!| \f\cdot \e)\leq -\log\l$, with $\l$ the Pimsner-Popa bound in \eqref{PP}, and in this case $\l^{-1}$ may be strictly less than Kosaki's index. 

Let $\A$ be a semifinite factor and $\B\subset \A$ a subfactor. If the index $\B\subset \A$ is finite (w.r.t. some expectation), then $\B$ is semifinite too. In this case, the trace $\tau$ of $\A$ has a semifinite restriction to $\B$ and  $[\A:\B]_\e <\infty$, with $\e$ the $\tau$-preserving expectation, see \eqref{tg} and\cite{L90}. 
\bcor\label{corO}
Let $\A$ be a semi-finite, infinite dimensional factor. If $\B\subset\A$ is a finite index subfactor and $\e:\A\to\B$ the expectation preserving the trace $\tau$, then
\[
\sup_\f\big\{S_\tau(\f|_\B) -S_\tau(\f)\big\}  = \log[\A:\B]_\e \, ,
\]
where the supremum is taken over all normal states $\f$ of $\A$ such that $S_\tau(\f)$ and $S_\tau(\f|_\B)$ are finite. \ecor
\proof
Suppose first that the trace $\tau$ is bounded. Then eq. \eqref{Se} in the proof of Prop. \ref{bound} shows that $S_\tau(\f|_\B) -S_\tau(\f)  = \log[\A:\B]_\e $
if $\f$ is a maximum point in eq. \eqref{max}. 

If $\tau$ is unbounded, we consider again a state $\f$ of $\A$ such that $S_\A(\f |\!| \f\cdot \e) = \log  [\A:\B]_\e $. 
We take a sequence of bounded entropy approximation functionals $\psi_k$ for $\tau$ w.r.t. $\f$ with $\psi_k\cdot\e = \psi_k$ as in the proof of Theorem \ref{bound2}. 
As $S(\f |\!| \psi_k)$ is finite, also $S(\f_\B |\!| \psi_k|_\B)$
is finite by Lemma \ref{finiteS}. Therefore $\psi_k|_\B$ is a sequence of bounded entropy approximation for $\tau|_\B$ w.r.t. $\psi_k|_\B$. 
By Lemma \ref{approx}, we so have
\[
S(\f |\!| \psi_k)\to S(\f |\!| \tau)\, ,\quad S(\f|_\B |\!| \psi_k|_\B)\to S(\f|_\B |\!| \tau|_\B)\, .
\]
By eq. \eqref{Petz}, we have
\[
S(\f |\!| \psi_k) - S(\f|_\B |\!| \psi_k|_\B) =
S(\f |\!| \psi_k\cdot\e) - S(\f|_\B |\!| \psi_k |_\B) =  S(\f |\!| \f\cdot\e) = \log[\A:\B]_\e \, ,
\]
thus
\[
S(\f |\!| \tau) - S(\f|_\B |\!| \tau|_\B) =  \lim_k \big( S(\f |\!| \psi_k) - S(\f|_\B |\!| \psi_k|_\B) \big)  =  \log[\A:\B]_\e \, ,
\]
that is $S_\tau(\f|_\B) -S_\tau(\f)  = \log[\A:\B]_\e$ due to the identity \eqref{Stau}. 
\endproof
More generally, let $\B\subset\A$ be an inclusion of von Neumann algebras with a finite index expectation $\e:\A\to\B$. We shall say that $\e$ has scalar index $ [\A:\B]_\e$ if eq. \eqref{Ki} holds for a scalar $ [\A:\B]_\e$. This is the case if the centers of $\A$ and $\B$ are finite dimensional and have trivial intersection, with $\e$ the minimal expectation \cite{GL}. 

The identity \eqref{SSI} still holds in this case, by the same proof. It follows that Corollary \ref{corO} remains true if $\B\subset\A$ is an inclusion of properly infinite von Neumann algebras which has finite scalar index $ [\A:\B]_\e$. 

\section{Further comments}
Structures in the physical literature (see \cite{GR,EW}) suggest to consider the entropy relative to a linear subspace, not only relative to an algebra. We consider such a notion and a few comments. 

Let $\A$ be a von Neumann algebra and $\f,\psi$ positive, normal, linear functionals on $\A$. Given a  linear subspace $V\subset \A$ containing the identity, we set
\ben
S_V(\f |\!|\psi) = \sup_{n\in\mathbb N}\sup_{x\in\mathfrak V}\left\{ \f(1)\log n - 
\int_{1/n}^\infty\big(\f(y(t)^*y(t)) + t^{-1}\psi(x(t)x(t)^*)\Big)\frac{dt}{t}\right\} \, ,
\een
where $\mathfrak V$ is the set of all step functions $x: (0, \infty)\to V$ with finite range, and $x(t) + y(t) =1$. 

If $V$ is $^*$strongly dense in $\A$, this is of course Kosaki's formula \eqref{Ko} for the relative entropy; namely
\[
S_V(\f |\!|\psi) = S_\A(\f |\!|\psi)\, .
\]
We list the following basic properties of $S_V$, whose proof is immediate. $\A$ is a von Neumann algebra, $\f,\psi,\phi$ normal, positive linear functionals on $\A$ and $V\subset\A$ a unital linear space. 
\begin{itemize}
\item[$a)$] $\phi \leq \psi$ implies $S_V(\f |\!| \phi) \geq S_V(\f |\!| \psi)$. 
\item[$b)$] $S_{\bar V}(\f |\!| \psi) = S_V(\f |\!| \psi)$ with $\bar V$ the $^*$strong closure of $V$. 
\item[$c)$] Monotonicity.  If $W\subset V$ is a unital linear subspaces, then $S_W(\f |\!| \psi)\leq  S_V(\f |\!| \psi)$ . 
\item[$d)$] Martingale convergence. Let $V_i\subset \A$ be an increasing net of unital linear subspaces with $V\equiv\cup_i V_i$. Then $S_{V_i}(\f |\!| \psi) \nearrow  S_V(\f |\!| \psi)$ . 
\end{itemize}
If now $\f$ is a positive, normal, linear functional and $\psi$ a normal, semifinite, faithful weight on $\A$, we set
\ben\label{SV}
S_V(\f |\!|\psi) = \inf_{\phi\leq\psi}S_V(\f |\!|\phi)
 \, ,
\een
where the infimum is taken over the set $\P_\psi$ of all positive, normal, linear functionals $\phi$ on $\A$ such that $\phi\leq \psi$. We recall that
\[
\psi(x) = \sup\big\{\phi(x) : \phi\in \P_\psi\big\}\, , \quad \text{for all positive}\ x\in\A\, ,
\]
\cite[Thm. 1.11]{T}. 

Suppose that $\tau$ is a semifinite, faithful normal trace on $\A$ and $\rho$ is the density matrix of $\f$ w.r.t. $\tau$. Recall that
\ben\label{SA}
S_\tau(\f) = - S_\A(\f |\!|\tau)\, ,
\een
provided the $S_\tau(\f) = -\tau(\rho\log\rho)$ is well defined. We may define $S_\tau(\f)$ for all states by  the above formula with the right hand side given by \eqref{SV} with $V =\A$. 

If $V\subset \A$ is a linear subspace containing the identity as above, we then set
\[
S_{\tau,V}(\f) \equiv - S_V(\f |\!|\tau) \,   .
\]
If $W\subset V$ is a unital linear subspace, it follows from $c)$ above that
\[
W \subset V \implies S_V(\f |\!|\tau) \leq S_W(\f |\!|\tau) \, ,
\]
therefore the monotonicity property holds for $S_{\tau,V}(\f)$:
\[
S_{\tau,V}(\f) \leq S_{\tau,W}(\f)  \, ,
\]
in particular $S_{\tau,\A}(\f) \leq S_{\tau,V}(\f)$.  
\bigskip

\noindent
{\bf Acknowledgements.} 
R.L.  acknowledges partial support by MIUR FARE R16X5RB55W QUEST-NET, GNAMPA-INdAM and
the MIUR Excellence Department Project awarded to the Department of Mathematics, University of Rome Tor Vergata, CUP E83C18000100006.
E.W. acknowledges NSF support under NSF-PHY1911298.

\end{document}